# Effects of Strain and Film Thickness on the Stability of the Rhombohedral Phase of HfO$_2$

Yuke Zhang[1], Qiong Yang[1,2,*], Lingling Tao[3], Evgeny Y. Tsymbal[3,†], and Vitaly Alexandrov[2,‡]

[1]*School of Materials Science and Engineering, Xiangtan University, Xiangtan, Hunan 411105, China*

[2]*Department of Chemical and Biomolecular Engineering, University of Nebraska, Lincoln, Nebraska 68588, USA*

[3]*Department of Physics and Astronomy, University of Nebraska, Lincoln, Nebraska 68588, USA*

The discovery of ferroelectric polarization in HfO$_2$-based ultrathin films has spawned a lot of interest due to their potential applications in data storage. Recently, a new *R*3*m* rhombohedral phase was proposed to be responsible for the emergence of ferroelectricity in the [111]-oriented Hf$_{0.5}$Zr$_{0.5}$O$_2$ thin films, but the fundamental mechanism of ferroelectric polarization in such films remains poorly understood. In this paper, we employ density-functional-theory calculations to investigate structural and polarization properties of the *R*3*m* HfO$_2$ phase. We find that the film thickness and in-plane compressive strain effects play a key role in stabilizing the *R*3*m* phase leading to robust ferroelectricity of [111]-oriented *R*3*m* HfO$_2$.

───────

[*] qyang@xtu.edu.cn

[†] tsymbal@unl.edu

[‡] valexandrov2@unl.edu

## I. INTRODUCTION

The HfO$_2$-based materials have aroused much interest in the last years due to their promise in nanoelectronics [1-5]. As compared with the conventional perovskite ferroelectrics, HfO$_2$-based ferroelectric thin films exhibit a range of properties appealing for low-power high-density memory applications [6,7], such as compatibility with the silicon-based semiconductor technology, robust switchable ferroelectricity at the nanoscale (<10 nm) and a large band gap of about 5-6 eV.

The atomistic origin for the emergence of switchable ferroelectric polarization in HfO$_2$-based ultrathin films, however, remains unclear [8]. The ground-state monoclinic $P2_1/c$ phase of HfO$_2$ is centrosymmetric and thus does not support ferroelectric polarization. Therefore, considerable prior efforts have been focused on identifying a ferroelectric HfO$_2$ phase that can be stable and displays pronounced ferroelectricity at the nanoscale as observed experimentally [3,8-11]. For example, by employing a group theoretical analysis in combination with density-functional-theory (DFT) calculations Huan *et al.* have previously identified two non-centrosymmetric polar orthorhombic phases with $Pca2_1$ and $Pmn2_1$ space groups that could be responsible for ferroelectricity in HfO$_2$-based films [8]. In particular, the $Pca2_1$ phase was most widely accepted as the ferroelectricity origin based on the numerous experimental and theoretical studies [3,8-11], which is also found to exhibit sizable spin-orbit coupling [12] and magnetoelectric effect at ferroelectric/ferromagnetic interface based on DFT calculations [13]. However, the phases $Pca2_1$ and $Pmn2_1$ are less stable thermodynamically than monoclinic $P2_1/c$ in the bulk. Though many factors such as strain (stress) [11,14], doping [15-17], defects [18], electric field [19] and surface energy [20], are found to be helpful for the stabilization of the orthorhombic $Pca2_1$ ferroelectric phase, a convinced and widely accepted mechanism for the stability of the possible ferroelectric phase of the HfO$_2$-based thin films is still ongoing.

Recently, Wei *et al*. has reported a rhombohedral ferroelectric phase with $R3m$ space group, which is different from previously described polar orthorhombic phases [21]. The $R3m$ phase was identified by using a combination of X-ray diffraction and transmission electron microscopy in the predominantly [111]-oriented Hf$_{0.5}$Zr$_{0.5}$O$_2$ thin film epitaxially grown on the La$_{0.7}$Sr$_{0.3}$MnO$_3$/SrTiO$_3$ substrate. One of the structural features of the observed phase is that the atomic plane spacing along the out-of-plane direction ($d_{111}$) is markedly larger than that along the [-111], [1-11] and [11-1] directions ($d_{-111}=d_{1-11}=d_{11-1}$) being consistent with a rhombohedral unit cell. The remnant polarization ($P_r$) was shown to reach a significant value of 34 μC/cm$^2$ for the Hf$_{0.5}$Zr$_{0.5}$O$_2$ film thickness of about 5 nm. However, according to the performed DFT calculations, the total energy of the found rhombohedral $R3m$ phase turns out to be even higher than that for orthorhombic $Pca2_1$. Moreover, a substantial strain was found to be required for HfO$_2$ (or Hf$_{0.5}$Zr$_{0.5}$O$_2$) to display ferroelectric polarization comparable with experimental values. It has been also detected experimentally that both the polarization magnitude and $d_{111}$-$d_{11\text{-}1}$ difference for the rhombohedral ferroelectric thin film increase upon decreasing the film thickness, which is important for the application of nanoscale HfO$_2$-based ferroelectrics. These experimental findings have stimulated our present study focusing on the behavior of the identified $R3m$ ferroelectric phase through the analysis of strain and size effects.

In this paper, we employ DFT calculations to provide further insights into the origin of ferroelectric polarization and phase stability of the $R3m$ ferroelectric polymorph of HfO$_2$. To this end, we focus on the experimentally relevant system of the [111]-oriented HfO$_2$ thin films and analyze both strain and size effects. We find that an in-plane compressive strain is essential for reproducing the experimentally observed $d_{111}$-$d_{11\text{-}1}$ difference and polarization enhancement in the [111]-oriented $R3m$ HfO$_2$ films. The size effect is also found to facilitate the phase stabilization under in-plane compressive strain of the film. The obtained results provide new insights into the emergence of stable ferroelectric polarization in HfO$_2$ thin films, which should be useful for a better understanding of the behavior of nanoscale ferroelectric films based on HfO$_2$.

## II. COMPUTATIONAL METHODOLOGY

In this work, we focus on the rhombohedral $R3m$, tetragonal $P4_2/nmc$, ferroelectric orthorhombic $Pca2_1$, and non-ferroelectric monoclinic $P2_1/c$ phases of HfO$_2$. All DFT calculations are performed using the projector augmented-wave (PAW) method as implemented in Vienna Ab initio Simulation Package (VASP) code [22]. The Perdew-Burke-Ernzerhof generalized gradient approximation (GGA-PBE) is used to describe exchange-correlation interactions [23]. The plane-wave cutoff energy is chosen to be 450 eV. The atomic coordinates of the $R3m$ phase are taken from Ref. [21]. The strain effect on ferroelectric polarization and phase stability of bulk HfO$_2$ is examined using the hexagonal cells (12 Hf and 24 O atoms) with their [0001] directions along the [111] directions of the conventional unit cells (4 Hf and 8 O atoms). The [111]-oriented periodic slabs are constructed with a vacuum gap of at least 18 Å in the out-of-plane direction. The 5×5×4 and 5×5×1 Monkhorst-Pack k-point meshes [24] are used for the bulk and thin-film calculations, respectively, to ensure the energy convergence. Structural optimizations are carried out until the Hellmann-Feynman forces on each atom are less than 10 meV/Å. The magnitude of ferroelectric polarization for the *R*3*m* phase is calculated within the Berry phase approach with taking the tetragonal *P*4$_2$/*nmc* phase as the non-polar reference state [25].

### III. RESULTS AND DISCUSSION

Figures 1(a) and 1(b) show the atomic structure and lattice vectors of the *R*3*m* rhombohedral phase of HfO$_2$ in its conventional unit cell (with 4 Hf and 8 O atoms). The hexagonal cell of rhombohedral *R*3*m* HfO$_2$ (with 12 Hf and 24 O atoms) which is employed to examine the strain effects is depicted in Fig. 1(c) with its out-of-plane direction along the [111] direction of the conventional unit cell (Figs. 1(a) and 1(b)). In our simulations the equiaxial strain was applied on the (001) plane of the hexagonal cell (i.e. the (111) plane of conventional cell as shown in Fig. 1(b)) with the out-of-plane lattice fully relaxed to stress-free to be consistent with the previous experimental work [21]. Figure 1(d) shows the change of the *d*-spacing as a function of the in-plane compressive strain. Under the strain-free conditions the conventional unit cell of *R*3*m* HfO$_2$ shown in Fig. 1(a) is cubic with $d_{111}$ being equal to $d_{11\text{-}1}$ (and $d_{\text{-}111}$=$d_{1\text{-}11}$=$d_{11\text{-}1}$). When the in-plane strain is increased, $d_{111}$ gradually goes up and $d_{11\text{-}1}$ goes down (see Fig. 1(d)). According to the scanning transmission electron microscopy (STEM) measurements, $d_{111}$ for the *R*3*m* Hf$_{0.5}$Zr$_{0.5}$O$_2$ phase ranges between 2.98 and 3.27 Å for the film thickness between 9 and 1.5 nm [21]. From the $d_{111}$-strain relation for HfO$_2$ (Fig. 1(d)) found in this work, the in-plane compressive strain is estimated to be between 1.5% and 5% to provide experimentally measured values of $d_{111}$.

Using this experimentally relevant range of compressive strains, we then calculate the out-of-plane spontaneous polarization (*P*) as plotted in Fig. 1(f). It is clearly seen that polarization is small for the in-plane strain below 3%, while it increases rapidly beyond 3% and reaches ~25 $\mu$C/cm$^2$ at the 5% strain. This result is in full agreement with the DFT computed polarization of the *R*3*m* phase of Hf$_{0.5}$Zr$_{0.5}$O$_2$ as a function of $d_{111}$ reported previously [21]. Thus, we conclude that a large in-plane strain is needed for the *R*3*m* phase HfO$_2$ to exhibit a pronounced ferroelectric polarization.

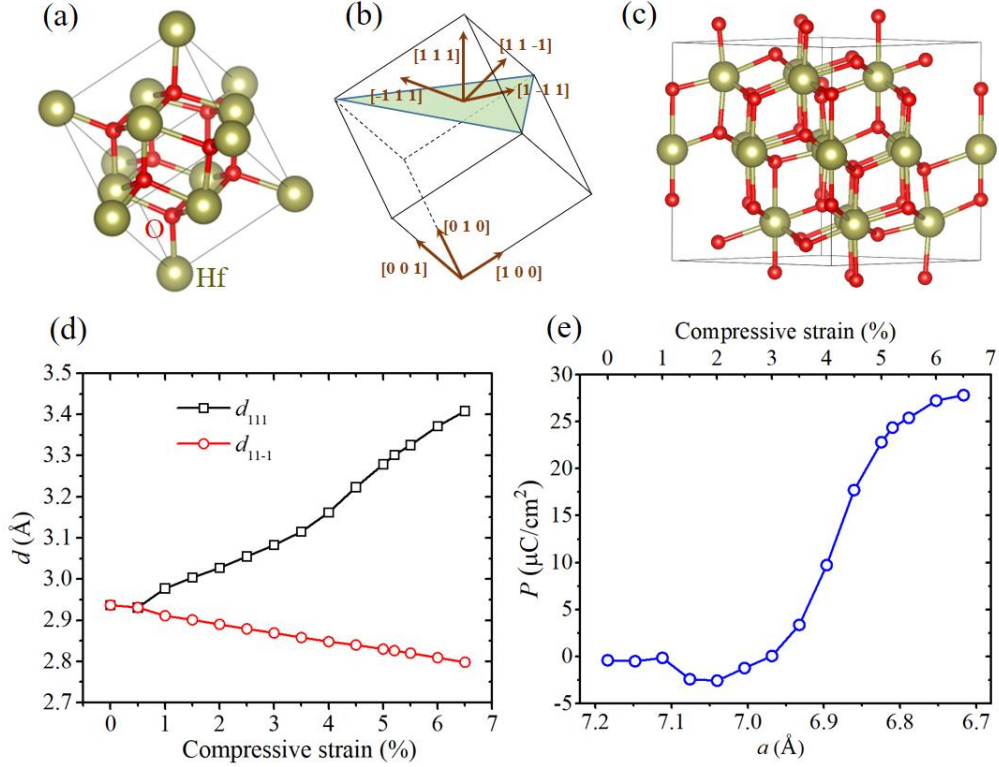

FIG. 1 The atomic structure (a) and lattice vectors (b) of $R3m$ phase $HfO_2$ in conventional cell. (c) The atomic structure of the [111]-oriented $R3m$ phase in hexagonal cell. (d) The $d$-spacing ($d$) of the $R3m$ phase $HfO_2$ versus in-plane compressive strain. (e) The spontaneous polarization ($P$) of the strained $R3m$ phase $HfO_2$. Upper and lower horizontal coordinates show the magnitudes of in-plane compressive strain and lattice constants, respectively.

We next provide insights into the origin of the increase of polarization as a function of strain for the $R3m$ phase of $HfO_2$. To this end, we first examine the atomic structures of the strain-free and compressively strained $R3m$ phase. For the strain-free case shown in Fig. 1(c), every O atom is coordinated with four Hf atoms forming an O-$Hf_4$ tetrahedron. This atomic structure is similar to that of the non-polar tetragonal $P4_2/nmc$ phase, and therefore the spontaneous polarization of the stress-free rhombohedral $R3m$ phase is negligible. For strained $R3m$ $HfO_2$ (Fig. 2), we observe that due to the in-plane compression and out-of-plane extension, the vertical Hf-O bonds become contracted and elongated, respectively, relative to the strain-free case. Specifically, by inspecting Hf-$O_2$ layers of the strained compound, we find that one of the four O atoms above Hf forms the O-$Hf_4$ tetrahedron with the upper Hf-$O_2$ layer, while three of the four O atoms below Hf form the O-$Hf_4$ tetrahedron with the lower Hf-$O_2$ layer. This bonding order causes the structural asymmetry and leads to polarization pointing upward in strained $R3m$ $HfO_2$.

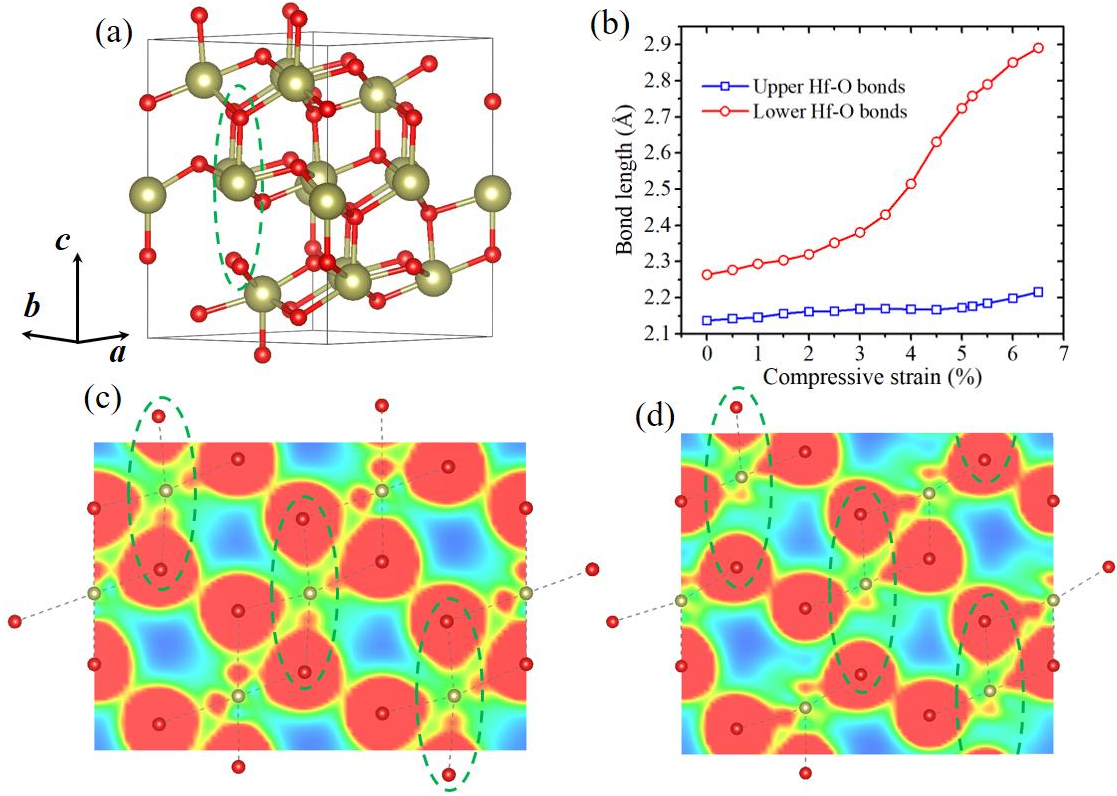

FIG. 2 (a) The atomic structure of the 6% compressively strained [111]-oriented $R3m$ HfO$_2$. (b) The length variation of the Hf-O bonds indicated by the dashed ellipse in (a). (c) and (d): The charge densities of the strain-free (c) and the compressively strained (d) $R3m$ HfO$_2$ along the (110) plane of the hexagonal cells.

The structural changes induced by the in-plane strain can be quantified by computing the bond length variations for the vertical Hf-O bonds, as denoted by the dashed ellipse in Fig. 2(a). Figure 2(b) shows the Hf-O bond length changes for the upper (shorter) and lower (longer) vertical Hf-O bonds upon increasing the in-plane compression. It is seen that the lengths for the short and long Hf-O bonds are almost the same in the strain-free condition. Under stress, the length of the short Hf-O bonds remains almost constant, whereas the length of the long Hf-O bonds increases substantially as the strain goes from 3% to 5%. This is consistent with the observed polarization enhancement under the in-plane strain shown in Fig. 1(e).

Next, we analyze the charge density redistribution under different in-plane strains. Figures 2(c) and 2(d) show the computed electron densities of strain-free and 6% compressively strained $R3m$ HfO$_2$, respectively, across the (110) plane of the hexagonal cell. It can be seen that in the 6% compressively strained case, the electron density between Hf and the lower O atoms marked by the dashed ellipses in Fig. 2(d) becomes depleted as compared to the strain-free case due to considerable elongation of these Hf-O bonds. Thus, it is determined that ferroelectric polarization of the compressively strained [111]-oriented $R3m$ HfO$_2$ phase originates from the disruption of Hf-O bonding order, which is similar to perovskite-structured ferroelectrics, such as BaTiO$_3$ and PbTiO$_3$ [26,27].

It is known from previous DFT calculations that the energy of $R3m$ HfO$_2$ lies above the ground-state $P2_1/c$ bulk phase by almost 160 meV/f.u [21]. It is also determined that the compressive strain required for pronounced ferroelectric polarization of [111]-oriented $R3m$ HfO$_2$ should be relatively high (~5%, Fig. 1(e)). Therefore, we next examine how stability of the bulk HfO$_2$ phases changes as a function of applied volumetric and in-plane stress. To this end, we consider tetragonal $P4_2/nmc$, orthorhombic ferroelectric $Pca2_1$ and the most-stable non-ferroelectric monoclinic $P2_1/c$ phases of HfO$_2$ and compare their stability with the $R3m$ phase. The volumetric stress is applied to the conventional cell of HfO$_2$ (with 4 Hf and 8 O atoms) by fixing the cell volume and fully relaxing the cell shape and atomic coordinates. Figure 3(a) shows the calculated phase stability diagrams relative to the ground-state $P2_1/c$ phase under volumetric stress. It is seen that regardless of the applied stress the bulk $R3m$ HfO$_2$ phase turns out to be the highest-energy polymorph.

We next analyze phase stability of the [111]-oriented HfO$_2$ films under the in-plane equiaxial strain (Fig. 3(b)). The in-plane strains are imposed on the (001) planes of the constructed hexagonal cells (i.e. the {111} planes of the conventional cells, see Fig. 1) of the considered $R3m$, $P4_2/nmc$, $Pca2_1$ and $P2_1/c$ phases with fixed lattices angles and fully relaxed out-of-plane lattice constants and all atomic positions. For each HfO$_2$ polymorph, there are four possible <111> crystal orientations ([111], [-111], [1-11] and [11-1] directions in conventional cells) that should be taken into consideration when investigating the effects of epitaxial strain.

The monoclinic $P2_1/c$ phase displays the same energy when the strain is applied in the planes perpendicular to the [111] and [1-11] directions, but exhibits different energies when the strain is applied in the planes perpendicular to the [-111] and [11-1] directions. Also, the orthorhombic $Pca2_1$ and tetragonal $P4_2/nmc$ phases are characterized by the same energies when the in-plane strains are applied perpendicular to any of the <111> directions. For the rhombohedral $R3m$ phase, we only plot results for the [111]-oriented case, since the strain imposed along the (11-1), (1-11) and (11-1) planes results in phase instability. Figure 3(b) shows the estimated energy variation upon changing the in-plane area of the hexagonal cells for the corresponding HfO$_2$ phases.

It is seen from the figure that the $P2_1/c$ phase remains the most stable and the $R3m$ phase is the least stable under the considered strains. We note here that Liu and Hanrahan have previously found that the $Pca2_1$ and $Pbca$ phases are more stable than the $P2_1/c$ phase for the [111]-oriented $HfO_2$ under the compressive strain [14]. The disagreement with our results could be because the difference of the in-plane areas of the $P2_1/c$ phase with different <111> orientations was not considered in their study.

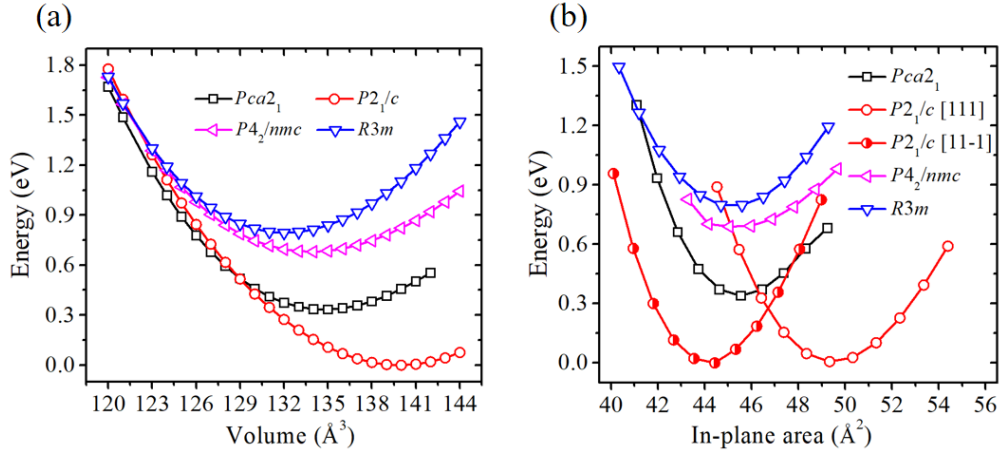

FIG. 3 (a) The total energies of different phases of $HfO_2$ in conventional cells (with 12 atoms) under volumetric strain. (b) The total energies of different phases of <111>-oriented $HfO_2$ (per 12 atoms) under plane strain.

From the above calculation, the $R3m$ phase of $HfO_2$ is of higher energy as compared with the other considered competing polymorphs under either volumetric strain or plane strain in the bulk form. Considering another experimentally observed fact that ferroelectric polarization of the $R3m$ $HfO_2$ film is more pronounced at smaller thicknesses [21], we further investigate the effect of thin-film size on the phase stability of [111]-oriented $HfO_2$. The <111>-oriented $HfO_2$ thin films are constructed with the Hf-$O_2$ layer numbers ($N$) varying from 1 to 9 as shown in Fig. 4(a). In these calculations, both the lattice and atomic coordinates are allowed to fully relax to obtain stress-free configurations of the $HfO_2$ thin films for each phase. Figure 4(b) shows the total energy of a thin film normalized per one Hf-$O_2$ layer (4 Hf and 8 O atoms) as a function of film thickness across the three $HfO_2$ polymorphs. Overall, it is seen that the energies of the $HfO_2$ thin films increase upon reducing film thickness for all phases due to the increasing proportion of surface energy. The energy of $P2_1/c$ phase increases rapidly when the film thickness decreases from 9 to 1 Hf-$O_2$ layers, while the variation of the energy for the $R3m$ phase is much slower. From the computed energies of thin-film and bulk $HfO_2$, the surface energies for $R3m$[111], $Pca2_1$[111], $P2_1/c$[111] and $P2_1/c$[11-1] are estimated at 2.48, 3.10, 3.65 and 2.79 eV per surface unit cell (i.e. 0.91, 1.10, 1.20 and 1.02 $J/m^2$), respectively, according to the formula (3) of Ref. [28]. Generally, these surface energies of {111} planes are much lower than those of the {100} planes in Ref. [20], which make the <111>-oriented epitaxial growth of $HfO_2$ films more favorable. It is interesting to notice in Fig. 4(b) that the $R3m$ phase is more stable than the $Pca2_1$ phase if the film thickness is less than 5 Hf-$O_2$ layers. Moreover, the $R3m$ $HfO_2$ becomes the most stable phase under extreme film thicknesses ($N = 1, 2$). This suggests that the $R3m$ $HfO_2$ phase may be stabilized under the small thickness due to its lower surface energy displaying robust ferroelectric polarization.

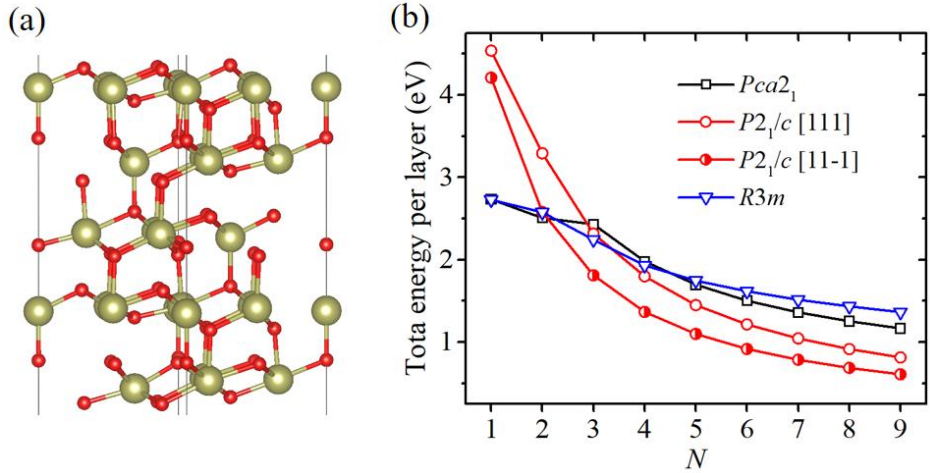

FIG. 4 (a) The atomic structure of the [111]-oriented $R3m$ phase $HfO_2$ thin film with $N$=5. (b) The total energies of different phases for the <111>-oriented $HfO_2$ thin films per Hf-$O_2$ layer. The total energy of the stress-free bulk $P2_1/c$ $HfO_2$ per 12 atoms are set to be zero. Since the $P4_2/nmc$ phase for many thicknesses are not stable, their energies are not given here. And the $Pca2_1$ phase is also not stable for $N$=1 and 2.

We next discuss structural properties of the <111>-oriented $HfO_2$ thin films. First, the $d$-spacing for the [111]-oriented $R3m$ $HfO_2$ thin films is analyzed for the strained and non-strained $R3m$ $HfO_2$ films. It is seen from Fig. 5(a) that $d_{111}$ is larger than $d_{11-1}$ for the $R3m$ $HfO_2$ films for all the thicknesses considered in this study. The difference between $d_{111}$ and $d_{11-1}$ gradually increases with the decrease of the film thickness, in agreement with experimental data obtained for $Hf_{0.5}Zr_{0.5}O_2$ [21]. Figure 5(b) shows how the thin-film energies and the in-plane lattice constants change when the <111>-oriented $HfO_2$ film thickness increases from $N = 1$ to 9. It is seen that the in-plane lattice constants decrease and the average energies increase upon reducing the film thickness. The decrease in the in-plane lattice constant of $R3m$ HfO$_2$ with $N$ is especially pronounced – it shrinks by almost 4% for the $N = 4$ film relative to bulk $R3m$ HfO$_2$. This indicates that the in-plane compressive strain for the $R3m$ HfO$_2$ thin film is substantially facilitated by the size effects. In Fig. 5(b), we also depict the difference between the two in-plane lattice constants for each phase through the error bars to show their in-plane structural asymmetry. It is seen that the two in-plane lattices are the same for the $R3m$ thin film, while the lengths for the two in-plane lattices for the [111]-oriented $Pca2_1$ and $P2_1/c$ phases are different. Furthermore, the difference in length for the two in-plane lattices for the [11-1]-oriented $P2_1/c$ phase is significant.

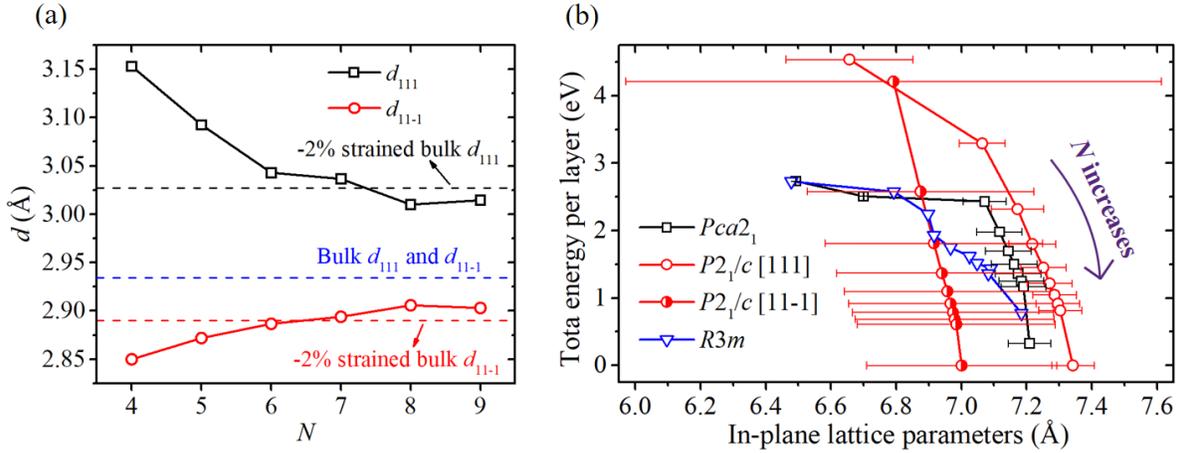

FIG. 5 (a) $d_{111}$ and $d_{11-1}$ of the $R3m$ HfO$_2$ thin films with different thicknesses. (b) Total energies per Hf-O$_2$ layer and in-plane lattice constants of the <111>-oriented HfO$_2$ thin films with different thicknesses. Symbols on each line from the left to right denote $N$=1 to 9, and the last point is the data of bulk for each phase. Error bars denote the length difference of the two in-plane lattices of each phase.

From the existing experiments and present calculations, it is seen that a substantial in-plane strain (~5%) is required for the [111]-oriented $R3m$ HfO$_2$ to exhibit a noticeable polarization. From Fig. 5(b), it is expected that the $R3m$ phase would be favorable if the (doped-) HfO$_2$ film is grown on a substrate with smaller in-plane lattice constants. Besides the epitaxial strain, the structural symmetry of the substrates is also believed to play an important role on the phase stability of the HfO$_2$ thin film from our calculation. The significant difference between the two in-plane lattice constants of the <111>-oriented $P2_1/c$ HfO$_2$ thin film (shown in Fig. 5(b)), especially the [11-1]-oriented one, indicates that the $P2_1/c$ phase may be inhibited when the (doped-) HfO$_2$ film is grown on a substrate with specific symmetry. Additionally, we believe that the interfacial effects and electrostatic screening from the electrode/$HfO_2$ interface could also be important factors for the stability of the *R*3*m* phase which have not been studied in this work, since the electrostatic screening would weaken the depolarization electric field in the thin film and make the polar phase, such as *R*3*m*, more stable. Furthermore, the effects of doping and defects may also play important roles in the stability and ferroelectric polarization of the *R*3*m* phase in practical $HfO_2$-based thin films.

## IV. CONCLUSIONS

In summary, we have examined the effects of strain and film thickness in the emergence of ferroelectricity in the rhombohedral phase-based $HfO_2$ thin films. Through the calculation of spontaneous polarization as a function of applied strain, we have demonstrated that relatively large strain (~5%) is necessary for the ferroelectric polarization of *R*3*m* phase to be comparable with the experimentally observed value. We have found that under the in-plane compressive strain, the [111]-oriented *R*3*m* phase exhibits a vertical long and short Hf-O bonding order, which is the origin for the significant ferroelectric polarization. However, the bulk *R*3*m* phase has been determined to be less stable than the ground-state $P2_1/c$ polymorph regardless of the imposed volumetric or plane strains. Nevertheless, it has been found that the *R*3*m* phase becomes the most stable phase at extremely small thickness of thin films. The obtained results suggest that the *R*3*m* phase responsible for the emergence of ferroelectricity in $HfO_2$ thin films could be stabilized under the combined effects of size and strain.

## ACKNOWLEDGMENTS

This work was supported by the National Natural Science Foundation of China (Grant No. 51702273), Research Foundation of Education Bureau of Hunan Province, China (Grant No. 18B056), National Science Foundation (NSF) through MRSEC (Grant No. DMR-1420645) and EPMD (Grant No. ECCS-1917635) Programs. Q. Yang thanks the China Scholarship Council (CSC) for financial support. The atomic structure was plotted using VESTA software.